\documentstyle[fleqn,twoside,gc-99]{article}

\begin{document}

\twocolumn[

\Arthead{0 (00)}{001}{003}

\Title{Dilatonic Black Holes Time Stability}

\Author{O.S. Khovanskaya\foom 1}
{Sternberg  Astronomical  Institute,   Universitetskii  Prospekt,  13,
Moscow 119992, RUSSIA}

\Rec{1 February 2002}

\Abstract
{The stability under small  time  perturbations of the dilatonic black
hole  solution  near the determinant curvature singularity is  proved.
This  fact  gives  the  additional  arguments  that  the  investigated
topological configuration can realise in nature. In the frames of this
model  primordial  black  hole  remnants are examined as  time  stable
objects, which  can form  an significant part of a  dark matter in the
Universe.}
]
\email 1 {khovansk@xray.sai.msu.ru}

One rather important  result of the  string gravity is  a  restriction
upon  the   minimal  black  hole  mass  \cite{alex97},  \cite{alex98},
\cite{khovansk} in  frames of the model  with the 4D  effective string
action,  containing  graviton,  dilaton  and higher order  cur  vature
corrections.  Let  me  consider  the  action  in  the  following  form
\cite{alex97}:
\begin{eqnarray}
\label{w}
S & = &  \frac{1}{16\pi} \int d^4 x \sqrt{-g} \biggl[
m^2_{Pl} \biggl( -R + 2 \partial_{\mu} \phi
\partial^\mu \phi \biggr)  \nonumber\\
& + &  \lambda \biggl( e^{-2\phi} (R_{ijkl}R^{ijkl}
- 4 R_{ij}R^{ij} \nonumber\\
& + & R^2) \biggr)  +   \ldots \biggr],
\end{eqnarray}
where  $m_{Pl}$  is  Plank  mass,  $\lambda$  is  the  string coupling
constant, $\phi$  is the dilatonic  scalar field. (The system of units
where $\hbar=c=G=1$ and  $m_{Pl}  = 1$ is used.  The  community of the
considered problem is not restricted by choosing $\lambda = 1$).

The  restriction  upon the minimal black  hole  mass is absent in  the
Schwarzschild   solution   of   Einstein's  classical  gravity.   This
"mathematical" result can be put into practice in  modern cosmology to
study  the   remnants   of   primordial   black  holes  \cite{mark66},
\cite{mark79}. Such remnants  can represent the final stage of Hawking
evaporation of primordial black  holes,  formed in the early Universe,
and  are   considered  as  dark  matter  candidates  \cite{khlop2000},
\cite{my}.

The search  of the  exact solutions or at least  the numerical ones in
the  offered  model  (\ref{w})   with   the  metric  depended  on  two
parameters, the  radial co-ordinate and the time, is  known to be very
difficult \cite{alex97}, \cite{alex98}, \cite{khovansk}. N evertheless
one can  receive the general  properties of the time-evolution of such
solutions  by  study   its   stability  about  time-parameter  in  all
particular points.

The black hole solution in the frames of considered model has only two
particular points. The first  one  is the usual coordinate singularity
$r_h$, which represents the event regular horizon of a black hole. The
second particular point is the  determinant  c  urvature  singularity,
which  has  the  infinite  derivatives   of   the   metric   functions
\cite{alex98}.

Primordial black hole  stability on the event horizon was investigated
in \cite{kanti}, \cite{jap}.

It was obtained  (P.  Kanti \cite{kanti}  at  al.) that the  dilatonic
black holes are stable near the black hole regular horizon $r_h$ under
linear time-dependent perturbations,  which  depend on only one radial
parameter. The metric parameterisation was
\begin{eqnarray*}
ds^2 = e^{\Gamma(r,t)}dt^2 - e^{\Lambda(r,t)}dr^2 - r^2 (d\, \theta^2 + \sin^2\, \theta d\, \psi^2)
\end{eqnarray*}
and the asymptotic forms of metric components and dilaton field near the regular horizon $r \approx r_h$ were
\begin{eqnarray*}
e^{-\Lambda(r)} & = & \lambda_1\,(r -  r_h) + \lambda_2\,(r - r_h)^{2} + \ldots, \\
e^{\Gamma(r)}   & = & \gamma_1\,(r -  r_h)  + \gamma_2\,(r - r_h)^{2} + \ldots, \\
\phi(r)& = & \phi_h + \phi_h'\,(r -  r_h) + \phi_h''\,(r - r_h)^{2} + \ldots,
\end{eqnarray*}
where $\lambda_1=2/(\lambda e^{\phi_h}\phi'_h/g^2+2r_h)$.

P. Kanti  at  al.  considered  perturbing  equations by time-dependent
linear perturbations of the form:
\begin{eqnarray*}
\Gamma(r,t)   & = & \Gamma(r) + \delta\,\Gamma(r,t) = \Gamma(r) + \delta\,\Gamma(r)\,e^{i\,\omega\,t}, \\
\Lambda(r,t)  & = & \Lambda(r) + \delta\,\Lambda(r,t) = \Lambda(r) + \delta\,\Lambda(r)\,e^{i\,\omega\,t}, \\
\phi(r,t)& = & \phi(r) + \delta\,\phi(r,t) = \phi(r) + \delta\,\phi(r)\,e^{i\,\omega\,t},
\end{eqnarray*}
where the  variations $\delta\,\Gamma(r,t), \delta\,\Lambda(r,t)$  and
$\delta\,\phi(r,t)$  were  assumed to be small. The stability  problem
was reduced to one-dimensional Schrodinger problem \cite{kanti}.

The regular  horizon stability was investigated  also in the  paper by
T.Torii  and  K.-i.Maeda \cite{jap}. They used the catastrophe  theory
and  compared  it  with  linear perturbation analysis.  Generally  the
catastrophe theory is a mathematical tool to investigate  a variety of
some physical states, T.Torii and K.-i.Maeda  showed  this  method  is
also  applicable  to  the  stability  analysis  of  various  types  of
non-Abelian black holes \cite{tach}.

It is necessary  to  make more  careful  stability analysis under  the
horizon with the  help  of suitable  choice  of asymptotic forms  that
approximate  the  metric  functions.  In this work I  investigate  the
dilatonic  black  hole  stability  near  the  determinant  curvat  ure
singularity $r=r_s$  \cite{alex98}.  After  the  definition  of a rest
point, this problem can be  reduced  to  a one-dimensional Schrodinger
problem under the variation of the field equation. One can  prove that
the small  time perturbations do not increase.  S o  in that case  the
solution of the dilatonic black hole is stable near $r_s$.

To investigate the stability under  time  per\-tur\-ba\-tions  of  the
dilatonic black  hole near the singularity $r \approx  r_s$, I use the
model (\ref{w})  with  the  following  non-static, asymptotically flat
spherically symmetric metric \cite{alex97}:
\begin{eqnarray}
\label{q}
ds^2 & = & \Delta dt^2 - {\displaystyle \frac {\sigma^2 } {\Delta}} dr^2 - r^2 (d\, \theta^2 + \sin^2\, \theta d\, \psi^2),
\end{eqnarray}
where $\Delta=\Delta(r,t)$, $\sigma=\sigma(r,t)$, $\phi=\phi(r,t)$.

According  to  Kanti's  method  \cite{kanti}  I   produce  the  metric
functions in the form:
\begin{eqnarray}
\label{ka1}
\Delta(r,t)=\Delta(r) + \delta\,\Delta(r,t) = \Delta(r) + \delta\,\Delta(r)\,e^{i\,\omega\,t},
\end{eqnarray}
\begin{eqnarray}
\sigma(r,t)=\sigma(r) + \delta\,\sigma(r,t) = \sigma(r) + \delta\,\sigma(r)\,e^{i\,\omega\,t},
\end{eqnarray}
\begin{eqnarray}
\label{ka2}
\phi(r,t)=\phi(r) + \delta\,\phi(r,t) = \phi(r) + \delta\,\phi(r)\,e^{i\,\omega\,t},
\end{eqnarray}
where  the  variations $\delta\,\Delta(r,t),  \delta\,\sigma(r,t)$ and
$\delta\,\phi(r,t)$ are assumed to  be  small \footnote{In this work I
only  study  the  small  time  perturbations  of  the  diagonal metric
components because in general I  investigate  the  case of spherically
symmetric metric. I'm  not interesting in some rotating effects, which
can appear because of non-zero non-diagonal metric components.}.

Thus   for   such   definition   of   variables   $r$   and   $t$   in
(\ref{ka1})-(\ref{ka2}) the  asymptotic forms of metric functions near
the singularity  $r \approx r_s$  depend on only radial coordinate $r$
and do not depend on time:
\begin{eqnarray}
\label{e}
\Delta (r) = \delta_0 + {\displaystyle \frac {\phi_2\,(r - r_s )}{\theta }} +  \delta_3\,(r - r_s )^{3/2} +  \ldots,
\end{eqnarray}
\begin{eqnarray}
\label{ba}
\sigma (r) = \sigma_0 + \sigma_2\,\sqrt{r - r_s } +  \ldots,
\end{eqnarray}
\begin{eqnarray}
\label{r}
\phi (r) =  \phi_0 + \phi_2\,(r - r_s ) +  \phi_3\,(r - r_s )^{3/2} + \ldots,
\end{eqnarray}
where
$$
\phi_0 = {\displaystyle \frac {3}{4}} \,{\rm ln}(2) -
{\displaystyle \frac {1}{2}} \,{\rm ln}( - \eta )
 + {\rm ln}(1 - \eta ) - {\rm ln}(r_s ),
$$
$$
\phi_2 = {\displaystyle \frac {1}{2}} \,{\displaystyle \frac {(
\eta  - 1)\,\sqrt{2}}{\eta \,r_s }},
$$
\begin{eqnarray*}
\phi_3 & = &  - \frac {1}{96 r_s \,\eta ^{4}} \, \sigma_2\,(\eta  - 1)^{2}\,\sqrt{2}\,( - 8\,\eta  + 8\, \eta ^{3} \\
 & - & 20\,\eta ^{2} + 4\,\eta ^{4} - 4\cdot 2^{3/4}\,\eta ^{2}
+ 3\cdot 2^{3/4}\,\eta ^{4} + 2^{3/4}),
\end{eqnarray*}
$$
\sigma_0 = 4\,{\displaystyle \frac {\eta }{1 - \eta ^{2}}},
$$
$$
\delta_0 =  - 16\,{\displaystyle \frac {\eta ^{2}}{(1 - \eta ^{2
})^{2}}},
$$
$$
\delta_3 =  - {\displaystyle \frac {16}{3}} \,{\displaystyle
\frac {\sigma_2\,(\eta  - 1)}{r_s \,(1 + \eta )^{2}}}
$$
\begin{center}
and
\end{center}
$$
\theta  = {\displaystyle \frac {1}{32}} \,{\displaystyle \frac {
\sqrt{2}\,(1 - \eta ^{2})^{2}}{\eta ^{2}}}.
$$

It is  possible to note  that $\eta  \in (-1,0)$. We  have three  free
parameters $\eta$, $\sigma_2$ and  $r_s$  \footnote{In this case it is
conveniently to choose $\eta$, $\sigma_2$ and $r_s$ as free. There are
direct three free parameters which  can  be reduced to the usual  free
parameters: the  black hole mass, the  dilaton charge and  the dilaton
value at infinity \cite{alex98}.}.

The exact field  equations for (\ref{w})-(\ref{q}) which depend on $r$
and $t$ are in Appendix.  Transforming  (\ref{t})-(\ref{y}),  one  can
obtain the  autonomous (over t) equation system of  the first order in
the following form:
\begin{equation}
\label{i}
\left\{
\begin{array}{rcl}
\dot\Delta & = & \alpha\\
\dot\sigma & = & \beta\\
\dot\phi & = & \gamma\\
\dot\alpha & = & -G \\
\dot\beta & = & 2G/\Lambda\\
\dot\gamma & = & F\\
\end{array}
\right.
\end{equation}
where $G,F$ are functions of  $r,  \Delta,  \Delta', \Delta'', \sigma,
\sigma', \phi, \phi', \phi''$ and $\alpha,\beta,\gamma$ are additional
variables.

The  function  $\Lambda = -2\Delta/\sigma$. Using the asimptotic  form
(\ref{e})-(\ref{ba}) one can obtained that $\Lambda=8 \eta/(1-\eta^2)$
for $r=r_s$. $\Lambda \ne 0, \Lambda \ne \infty$  near the singularity
$r \approx  r_s$ because the  dilaton function $ \phi(r)$ (\ref{r}) is
limited in this region \cite{alex98}.

Let  me  examine  the  equilibrium  states  of  the  autonomous system
(\ref{i}).

Let the point  $  (\Delta^*, \, \sigma^*, \,  \phi^*,  \, \alpha^*, \,
\beta^*, \, \gamma^*) $ is  some  rest point of the system  (\ref{i}),
that is for
$$
f_i \in \{ \alpha, \, \beta, \, \gamma, \, -G, \, 2G/\Lambda, \, F \},
$$
$$i=1,2, \ldots ,6$$
the condition
$$
f_i(\Delta^*, \, \sigma^*, \, \phi^* \, \alpha^*, \, \beta^*, \, \gamma^*)=0
$$
is executed. The trivial "equilibrium-like" solution which corresponds
to the given rest point  is  asymptotically stable if the first  order
system  is   stable.  It  occurs  when  the  all   roots  $s$  of  the
characteristic equation
\begin{eqnarray}
\label{n}
det \Bigl[ \Bigl(\frac{df_i}{dy_k} \Bigr)|_{rest\,\,point } - s\cdot \delta^i_k \Bigl] & = & 0
\end{eqnarray}
where $y_k \in \{\Delta^*, \, \sigma^*, \, \phi^*, \, \alpha^*, \, \beta^*, \, \gamma^* \}$ ($k=1,2, \ldots ,6$) have the negative real parts.

It is possible to find the rest point of the system (\ref{i}) using the asymptotic forms (\ref{e})-(\ref{r}) near the singularity $r \approx r_s$ from the condition:
\begin{equation}
\label{j}
\left\{
\begin{array}{rcl}
F|_{rest\,point} & = & 0\\
G|_{rest\,point} & = & 0\\
\end{array}
\right.
\end{equation}
In  the rest point  $  \alpha=\beta=\gamma  \equiv  0 $  and  $\eta  =
\eta^*=C_1, \,\, \sigma_2=\sigma_2(r_s)=C_2/r_s^{1/2}<0 $, where $C_1$
and $C_2$ are the number coefficients.

By solving (\ref{n})  one can receive  the pure imaginary  roots  $s$.
Thus  some  additional investigations for determine the stability  are
required.

Using  the  method of  linear  stability  one  can  rewrite  the field
equations (\ref{t})-(\ref{y}) with the variations:
\begin{eqnarray}
\label{kl}
0 & = & A_{i\,1}\delta\,\ddot \phi + A_{i\,2}\delta\,\dot \phi + A_{i\,3}\delta\,\phi +  A_{i\,4}\delta\,\phi' \nonumber\\
  & + & A_{i\,5}\delta\,\phi''  +  B_{i\,1}\delta\,\ddot \Delta + B_{i\,2}\delta\,\dot \Delta + B_{i\,3}\delta\,\Delta  \nonumber\\
  & + & B_{i\,4}\delta\,\Delta' + B_{i\,5}\delta\,\Delta''  +  C_{i\,1}\delta\,\ddot \sigma + C_{i\,2}\delta\,\dot \sigma \nonumber\\
  & + & C_{i\,3}\delta\,\sigma +  C_{i\,4}\delta\,\sigma' + C_{i\,5}\delta\,\sigma'',
\end{eqnarray}
where $i=1..4$.
Nonzero coefficients are in Appendix.

In  the  vicinity  of  finding  from  (\ref{j})  rest  point $O^{+}(0,
(r-r_s))$,  $(r-r_s)  \sim  10^{-3}$  it is possible to  evaluate  the
coefficients  in  front  of  the  variations  in (\ref{kl}) using  the
asymptotic  forms  (\ref{e})-(\ref{r}).  The simplified equation  from
(\ref{kl}) becomes the following:

\begin{eqnarray}
\label{m}
A(\delta\phi)'' + B(\delta\phi)' + C(\delta\phi) & = & \omega^2(\delta\phi),
\end{eqnarray}
where $A$,$B$ and $C$ near the singularity $r \approx r_s$ are:
\begin{eqnarray*}
A & = & \frac{a_1}{r_s} + \frac{a_2\,(r-r_s)^{1/2}}{r_s^{3/2}} + O(r-r_s),\\
B & = & \frac{b_1}{r_s^2} + \frac{b_2\,(r-r_s)^{-1/2}}{r_s^{5/2}} + O((r-r_s)^{1/2}), \\
C & = & \frac{c_1}{r_s^3} + \frac{c_2\,(r-r_s)^{-1/2}}{r_s^{5/2}} + O((r-r_s)^{1/2}),
\end{eqnarray*}
where $a_1,a_2,b_1,b_2,c_1,c_2$  depend  on $\sigma_2$ and $\eta$. For
$r=r_s$ these coefficients are number constants.

The eigenvalue $\omega^2$ in (\ref{m}) has only positive values. Thus,
the small time perturbations do not increase (they  can oscillate). So
in that  case the  solution is stable in the  vicinity of finding rest
point $(O^{+}(0,(r-r_s))$.

Taking into account both results: the stability of  the solution under
time perturbations at the regular event horizon $r_h$ \cite{kanti} and
at  the  determinant curvature singularity $r_s$, it  is  possible  to
conclude that the solution  of dilatonic black ho le is stable  in all
particular points. In application  to  cosmology this fact can confirm
the existence of remnants of primordial black holes,  which are stable
during time evolution.

I would like to gratefully  acknowledge  to Prof. M.V. Sazhin and  Dr.
S.O. Alexeyev for useful comments.

\vskip100mm

{\Large{Appendix}}

\small

The exact field  equations for (\ref{w})-(\ref{q}) which depend on $r$
and $t$ are the following unwieldy form:
\begin{eqnarray}
\label{t}
0 & = & 2\,e^{2\,\phi}\,r^{2}\,{\phi'}^{2}\,\sigma^{3}\,\Delta^{2}  -          8\,\phi''\,\sigma^{3}\,\Delta^{2} \nonumber\\
  & - & 2\,e^{2\,\phi}\,r\,\sigma'\,\sigma^{2}\,\Delta^{2} +                  16\,\phi'^{2}\,\sigma^{3}\,\Delta^{2} \nonumber\\
  & + & 8\,\phi'\,\sigma'\,\sigma^{2}\,\Delta^{2} +
        2\,\sigma^{5}\,e^{2\,\phi}\,r^{2}\,\dot \phi^{2} \nonumber\\
  & + & 8\,\dot \phi\,\dot \sigma\,\sigma^{4} -                                                     8\,\dot \phi\,\dot \sigma\,\sigma^{2}\,\Delta  \nonumber\\
  & - & 16\,\dot \phi^{2}\,\sigma^{3}\,\Delta +                                               16\,\sigma^{5}\, \dot \phi^{2} \nonumber\\
  & - & 8\,\sigma ^{5}\,\ddot \phi +
        8\,\ddot \phi\,\sigma^{3}\,\Delta \nonumber\\
  & - & 16\,\Delta^{3}\,\phi'^{2}\, \sigma
        - 24\,\Delta ^{3}\,\phi'\,\sigma' +  8\,\Delta ^{3}\,\phi''\,\sigma,
\end{eqnarray}
\begin{eqnarray}
0 & = &  - 2\,e^{2\,\phi}\,\sigma^{4}\,\Delta^{2}+            2\,e^{2\,\phi}\,r\,\Delta'\,\sigma^{2}\,\Delta^{2} \nonumber\\
  & - & 2\,\Delta^{3}\,e^{2\,\phi}\,r^{2}\,\phi'^{2}\,\sigma^{2}
         +  24\,\Delta^{3}\,\phi'\,\Delta' +  8\,\dot \phi \,\dot \Delta \,\sigma^{2}\,\Delta          \nonumber\\
  & - & 8\,\phi'\,\Delta'\,\sigma^{2}\,\Delta ^{2} +  2\,\Delta ^{3}\,e^{2\,\phi}\,\sigma^{2}          \nonumber\\
  & - & 16\,\ddot \phi \,\sigma \,\Delta^{2} +  32\,\dot \phi ^{2}\,\sigma ^{2}\,\Delta ^{2}          \nonumber\\
  & - & 32\,\dot \phi ^{2}\,\sigma^{4}\,\Delta  +  16\,\ddot \phi\,\sigma ^{4}\,\Delta  - 8\,\dot \phi\,\dot \Delta \,\sigma ^{4} \nonumber\\
  & - & 2\,e^{2\,\phi}\,r^{2}\,\dot \phi ^{2}\,\sigma ^{4}\,\Delta,
\end{eqnarray}
\begin{eqnarray}
0 & = &  - 8\,\dot \Delta ^{2}\,\sigma ^{3}\,\Delta  - 8\,\Delta''\,\sigma^{3}\,\Delta ^{3} - 16\,\ddot \sigma \,\sigma ^{2}\,\Delta ^{3} \nonumber\\
  & - & 8\,\dot \Delta \,\dot \sigma \,\sigma ^{2}\,\Delta ^{2} +  4\,\sigma ^{5}\,e^{2\,\phi }\,r
^{2}\,\ddot \phi \,\Delta ^{2} + 8\,\Delta ^{4}\,\Delta''\,\sigma \nonumber\\
  & - & 8\,\Delta ^{4}\,e^{2\,\phi }\,r\,\phi'\,\sigma ^{3} - 24\,\Delta ^{4}\,\Delta'\,\sigma'  + 8\,\Delta'\,\sigma'\,\sigma ^{2}\,\Delta ^{3} \nonumber\\
  & + & 32\,\dot \sigma ^{2}\,\sigma \,\Delta ^{3}  + 8\,\ddot \Delta \,\sigma ^{3}\,\Delta ^{2} + 8\,\Delta'^{2}\,\sigma \,\Delta ^{3} \nonumber\\
  & - & 8\,\sigma ^{5}\,\ddot \Delta \,\Delta  + 16\,\sigma^{5}\,\dot \Delta ^{2}                 - 24\,\dot \Delta \,\dot \sigma \,\sigma
^{4}\,\Delta  \nonumber\\
  & + & 16\,\ddot \sigma \,\sigma ^{4}\,\Delta ^{2} -  4\,\Delta ^{4}\,e^{2\,\phi}\,r
^{2}\,\phi''\,\sigma ^{3} \nonumber\\
  & - & 4\,e^{2\,\phi}\,\Delta'\,r^{2}\,\phi'\,\sigma ^{3}\,\Delta ^{3}  + 4\,\Delta ^{4}\,e^{2\,\phi}\,r^{2}\,\phi'\,\sigma'\,\sigma ^{2} \nonumber\\
  & + & 4\,e^{2\,\phi}\,\dot \sigma \,r^{2}\,\dot \phi \,\sigma ^{4}\,\Delta ^{2}  -  4\,\sigma ^{5}\,e^{2\,\phi}\,r
^{2}\,\dot \phi \,\dot \Delta \,\Delta,
\end{eqnarray}
\begin{eqnarray}
\label{y}
0 & = & 2\,e^{2\,\phi}\,r\,\Delta''\,\Delta^{3}\,\sigma ^{3} + 128\,\dot \phi \,\dot\sigma \,\phi'\,\Delta ^{3}\,\sigma ^{2} +  16\,\phi'\,\ddot\Delta \,\Delta ^{2}\,\sigma ^{3} \nonumber\\
  & - & 32\,\Delta ^{4}\,\phi'^{2}\,\Delta'\,\sigma + 32\,\dot\phi ^{2}\,\Delta'\,
\Delta ^{2}\,\sigma ^{3} + 32\,\dot\phi' \,\dot\Delta \,\Delta ^{2}\,
\sigma ^{3} \nonumber\\
  & + & 16\,\dot\phi \,\Delta'\,\dot\sigma \,\Delta ^{2}\,
\sigma ^{2} - 64\,\dot\phi\, \dot\Delta \phi'\,\Delta ^{2}\,
\sigma ^{3} +  16\,\Delta ^{4}\,\phi'\,\Delta'' \,\sigma \nonumber\\
  & - & 16\,\ddot\phi \,\Delta'\,\Delta ^{2}\,\sigma ^{3} - 16\,\phi'\,\dot\Delta ^{2}\,
\Delta \,\sigma ^{3} -  2\,e^{2\,\phi}\,r\,\Delta'\,\sigma'\,\Delta ^{3}\,\sigma ^{2} \nonumber\\
  & - & 4\,\sigma ^{5}\,e^{2\,\phi }\,r\,\dot\Delta ^{2} - 4\,e^{2\,\phi }\,r\,\ddot\sigma \,\Delta ^{2}\,\sigma ^{4} - 48\,\Delta ^{4}\,\phi'\,\Delta'\,\sigma' \nonumber \\
  & + & 32\,\ddot\phi\,\sigma'\,\Delta ^{3}\,\sigma ^{2} + 16\,\Delta ^{4
}\,\phi'' \,\Delta'\,\sigma - 16\,\dot\phi \dot\Delta \,\sigma'\,\Delta ^{2}\,
\sigma ^{2} \nonumber\\
  & - & 4\,\sigma ^{5}\,e^{2\,\phi}\,r\,\dot\phi ^{2}\,\Delta ^{2} + 64\,\phi'\,\dot\sigma ^{2}\,\Delta^{3}\,\sigma  + 4\,e^{2\,\phi}\,\Delta'\,\Delta ^{3}\,\sigma ^{3} \nonumber\\
  & - & 64\,\dot\phi ^{2}\,\sigma'\,\Delta ^{3}\,\sigma ^{2} -  64\,\dot\phi'\,\dot\sigma \,\Delta ^{3}\,\sigma ^{2} + 16\,\Delta'^{2}\,\phi'\,\Delta ^{3}\,\sigma \nonumber\\
  & - & 32\,\phi'\,\ddot\sigma \,\Delta ^{3}\,\sigma ^{2}  -  16\,\phi'\,\dot\Delta \,\dot\sigma \,\Delta ^{2}\,
\sigma ^{2} \nonumber\\
  & + & 6\,e^{2\,\phi}\,\dot\sigma \,r\,\Delta'\,\Delta \,\sigma ^{4}  + 2\,\sigma ^{5}\,e^{2\,\phi}\,r\,\ddot\Delta \,\Delta  - 4\,\Delta ^{4}\,e^{2\,\phi}
\,\sigma'\,\sigma ^{2} \nonumber\\
  & + & 4\,\Delta ^{4}\,e^{2\,\phi}\,r\,\phi'^{2}\,\sigma ^{3},
\end{eqnarray}

Nonzero coefficients for the field equations (\ref{t})-(\ref{y}) with the variations (\ref{kl}):
\begin{eqnarray*}
{A_{4 \,1}} & = &  - 2\,\Delta '\,\sigma ^{3}\,\Delta ^{2} + 4\,\sigma '\,\sigma ^{2}\,\Delta ^{3}, \\
{A_{4 \,3}} & = & \Delta ^{4}\,r\,\phi '^{2}\,\sigma ^{3}\,e^{2\,\phi } +
\Delta '\,\sigma ^{3}\,e
^{2\,\phi }\,\Delta ^{3} - \Delta ^{4}\,\sigma '\,\sigma ^{2}\,e^{2\,\phi } \\
& + & {\displaystyle
\frac {1}{2}} \,r\,\Delta ''\,\sigma ^{3}\,e^{2\,\phi }\,\Delta ^{3} -  {\displaystyle \frac {1}{2}} \,r\,\Delta '\,\sigma '\,\sigma ^{2}\,e^{2\,\phi }\,
\Delta ^{3}, \\
{A_{4 \,4}} & = & \Delta ^{4}\,r\,\phi '\,\sigma ^{3}\,e^{2\,\phi } - 6\,
\Delta ^{4}\,\Delta '\,\sigma ' + 2\,\Delta ^{4}
\,\Delta ''\,\sigma \\
 & - & 8\,\Delta ^{4}\,\phi '\,\Delta '\,\sigma  +  2\,\Delta '^{2}\,\sigma \,\Delta ^{3}, \\
{A_{4 \,5}} & = & 2\,\Delta ^{4}\,\Delta '\,\sigma,\\
{B_{4 \,1}} & = & 2\,\phi '\,
\sigma ^{3}\,\Delta ^{2} + {\displaystyle \frac {1}{4}} \,
\sigma ^{5}\,r\,e^{2\,\phi }\,\Delta ,\\
{B_{4 \,3}} & = &  - 24\,\Delta ^{3}\,\phi '\,\Delta '\,\sigma ' + 8\,\Delta ^{3}\,\phi ''\,\Delta '\,\sigma  \\
 & + & 6\,\Delta '^{2}\,\phi '\,\sigma \,\Delta ^{2} - 16\,\Delta ^{3}\,\phi '^{2}\,\Delta '\,\sigma  \\
 & + & 2\,\Delta ^{3}\,r\,\phi '^{2}\,\sigma ^{3}\,e^{2\,\phi }  -  2\,\Delta ^{3}\,\sigma '\,\sigma ^{2}\,e^{2\,\phi } \\
 & - & {\displaystyle \frac {3}{4}} \,r\,     \Delta '\,\sigma '\,\sigma ^{2}\,e^{2\,\phi }\,
\Delta ^{2}  +  {\displaystyle \frac {3}{4}} \,r\,\Delta ''\,\sigma ^{3}\,e^{
2\,\phi }\,\Delta ^{2}  \\
 & + & {\displaystyle \frac {3}{2}} \,\Delta '\,\sigma ^{3}\,e^{2\,\phi }\,
\Delta ^{2} + 8\,\Delta ^{3}\,\phi '\,\Delta''\,\sigma ,\\
{B_{4 \,4}} & = & 2\,\Delta ^{4}\,\phi ''\,\sigma  - 6\,\Delta ^{4}\,\phi '\,\sigma ' - {\displaystyle \frac {1}{4}} \,r\,\sigma '\,\sigma ^{2}\,e^{2\,\phi
}\,\Delta ^{3} \\
 & - & 4\,\Delta ^{4}\,\phi '^{2}\,\sigma  + {\displaystyle \frac {1
}{2}} \,\sigma ^{3}\,e^{2\,\phi }\,\Delta ^{3}  +  4\,\Delta '\,
\phi '\,\sigma \,\Delta ^{3},\\
{B_{4 \,5}} & = & 2\,\Delta ^{4}\,\phi '\,\sigma  + {\displaystyle \frac {1}{4}} \,r\,
\sigma ^{3}\,e^{2\,\phi }\,\Delta ^{3},\\
{C_{4 \,1}} & = &  - 4\,\phi '\,\sigma ^{2}\,\Delta ^{3} - {\displaystyle \frac {1}{2}}
\,r\,\sigma ^{4}\,e^{2\,\phi }\,\Delta ^{2},\\
{C_{4 \,3}} & = &  - {\displaystyle \frac {1}{2}} \,r\,\Delta '\,\sigma '\,\sigma \,e^{2\,\phi }\,
\Delta ^{3} + {\displaystyle \frac {3}{2}} \,\Delta ^{4}\,r
\,\phi '^{2}\,\sigma ^{2}
\,e^{2\,\phi } \\
 & + & {\displaystyle \frac {3}{2}} \,\Delta '\,\sigma ^{2}\,e^{2\,\phi }\,
\Delta ^{3}  -  \Delta ^{4}\,\sigma '\,\sigma \,e^{2\,\phi } \\
 & + & 2\,\Delta ^{4}\,\phi '\,\Delta '' - 4\,\Delta ^{4}\,\phi '^{2}\,\Delta ' + 2\,\Delta '^{2}\,\phi '\,\Delta ^{3} \\
 & + & 2\,\Delta ^{4}\,\phi ''\,\Delta ' + {\displaystyle \frac {3}{4}} \,r\,\Delta ''\,\sigma ^{2}\,e^{2\,\phi }\,\Delta ^{3},\\
{C_{4 \,4}} & = & - {\displaystyle \frac {1}{4}} \,r\,\Delta '\,\sigma ^{2}\,e^{2\,\phi }\,
\Delta ^{3} - {\displaystyle \frac {1}{2}} \,\Delta ^{4}\,
\sigma ^{2}\,e^{2\,\phi } - 6\,\Delta ^{4}\,\phi '\,\Delta ',\\
{A_{3 \,1}} & = & {\displaystyle \frac {1}{2}} \,\sigma ^{5}\,r
^{2}\,e^{2\,\phi }\,\Delta ^{2},\\
{A_{3 \,3}} & = & - \Delta ^{4}\,r^{2}\,\phi ''\,\sigma ^{3}\,e^{2
\,\phi } - 2\,\Delta ^{4}\,r\,\phi '\,\sigma ^{3}\,e^{2\,\phi } \\
& + & \Delta ^{4}\,r^{2}\,\phi '\,\sigma '\,\sigma ^{2}\,e^{2\,\phi } - \Delta '\,r^{2}\,\phi '\,\sigma ^{3}\,e^{2\,\phi }\,\Delta ^{3},\\
{A_{3 \,4}} & = &  - \Delta ^{4}\,r\,\sigma ^{3}\,e
^{2\,\phi } + {\displaystyle \frac {1}{2}} \,\Delta ^{4}
\,r^{2}\,\sigma '\,\sigma ^{2}\,e^{2\,\phi } \\
& - & {\displaystyle \frac {1}{2}} \,\Delta '\,r^{2}\,\sigma ^{3}\,e^{2\,\phi }\,\Delta ^{3},\\
{A_{3 \,5}} & = &  - {\displaystyle \frac {1}{2}} \,\Delta ^{4}
\,r^{2}\,\sigma ^{3}\,e^{2\,\phi} ,\\
{B_{3 \,1}} & = & \sigma ^{3}\,\Delta ^{2} - \sigma ^{5}\,
\Delta ,\\
{B_{3 \,3}} & = & 3\,\Delta '^{2}\,\sigma \,\Delta ^{2} - 4\,\Delta ^{3}\,
r\,\phi '\,\sigma ^{3}\,e
^{2\,\phi } - 2\,\Delta ^{3}\,r^{2}\,\phi ''\,\sigma ^{3}\,e^{2\,\phi } \\
 & + & 4\,\Delta ^{3}\,\Delta ''\,\sigma  +  3\,\Delta '\,\sigma '\,\sigma ^{2}\,
\Delta ^{2} - 12\,\Delta ^{3}\,\Delta '\,\sigma ' \\
 & - & 3\,\Delta''\,\sigma ^{3}\,\Delta ^{2}  -  {\displaystyle \frac {3}{2}} \,\Delta '\,r^{2}\,\phi '\,\sigma ^{3}\,e^{2\,\phi }\,\Delta ^{2} \\
 & + & 2\,\Delta ^{3}\,r^{2}\,\phi '\,\sigma '\,\sigma^{2}\,e^{2\,\phi },\\
{B_{3 \,4}} & = &  - 3\,\Delta ^{4}\,\sigma ' - {\displaystyle \frac {1}{2}} \,r^{2
}\,\phi '\,\sigma ^{3}\,e
^{2\,\phi }\,\Delta ^{3} + \sigma '\,\sigma ^{2}\,\Delta ^{3} \\
& + & 2\,\Delta '\,
\sigma \,\Delta ^{3},\\
{B_{3 \,5}} & = & \Delta ^{4}\,\sigma  - \sigma ^{3}\,
\Delta ^{3},\\
{C_{3 \,1}} & = &  - 2\,\sigma ^{2}\,\Delta ^{3} + 2\,\sigma
^{4}\,\Delta ^{2},\\
{C_{3 \,3}} & = & \Delta'^{2}\,\Delta ^{3} - 3\,\Delta ^{4}\,r\,\phi '\,\sigma ^{2}\,e^{2\,\phi }  - {\displaystyle \frac {3}{2}} \,\Delta ^{4}\,r^{
2}\,\phi ''\,\sigma ^{2}\,e^{2\,\phi } \\
 & + & 2\,\Delta '\,\sigma '\,\sigma \,\Delta^{3} -  3\,\Delta ''\,\sigma ^{2}\,\Delta ^{3} \\
 & - & {\displaystyle\frac {3}{2}} \,\Delta '\,r^{2}\,\phi '\,\sigma ^{2}
\,e^{2\,\phi (r)}\,\Delta ^{3}  +  \Delta ^{4}\,\Delta '' \\
 & + & \Delta ^{4}\,r^{2}\,\phi '\,\sigma'\,\sigma \,e^{2\,\phi },\\
{C_{3 \,4}} & = & \Delta '\,
\sigma ^{2}\,\Delta ^{3} - 3\,\Delta ^{4}\,\Delta ' + {\displaystyle \frac {1}{2
}} \,\Delta ^{4}\,r^{2}\,\phi '\,\sigma ^{2}\,e^{2\,\phi },\\
{A_{2 \,1}} & = & 2\,\sigma ^{4}\,\Delta  - 2\,\sigma ^{2}
\,\Delta ^{2},\\
{A_{2 \,3}} & = & {\displaystyle \frac {1}{2}} \,r\,\Delta '\,\sigma ^{2}\,e^{
2\,\phi }\,\Delta ^{2} - {\displaystyle \frac {1}{2}} \,
\sigma ^{4}\,e^{2\,\phi }\,\Delta ^{2} \\
& + &
{\displaystyle \frac {1}{2}} \,\Delta ^{3}\,\sigma ^{2}\,e
^{2\,\phi } - {\displaystyle \frac {1}{2}} \,\Delta ^{3}\,r^{
2}\,\phi '^{2}\,\sigma ^{
2}\,e^{2\,\phi },\\
{A_{2 \,4}} & = &  - {\displaystyle \frac {1}{2}} \,\Delta ^{3}
\,r^{2}\,\phi '\,\sigma
^{2}\,e^{2\,\phi } + 3\,\Delta ^{3}\,\Delta ' - \Delta '\,\sigma '^{2}\,\Delta '^{2},\\
{B_{2 \,3}} & = &  - {\displaystyle \frac {3}{4}} \,\Delta
^{2}\,r^{2}\,\phi '^{2}
\,\sigma ^{2}\,e^{2\,\phi } - {\displaystyle \frac {1}{2}
} \,\sigma ^{4}\,e^{2\,\phi }\,\Delta \\
& + & {\displaystyle \frac {1}{2}} \,r\,\Delta '\,\sigma ^{2}\,e^{2\,\phi
}\,\Delta  - 2\,\phi '\,\Delta '\,\sigma ^{2}
\,\Delta  + {\displaystyle \frac {3}{4}} \,\Delta ^{2}\,
\sigma ^{2}\,e^{2\,\phi }  \\
& + & 9\,\Delta ^{2}\,\phi '\,\Delta ',\\
{B_{2 \,4}} & = & 3\,\Delta ^{3}\,\phi ' - \phi '\,
\sigma ^{2}\,\Delta ^{2} + {\displaystyle \frac {1}{4}} \,r
\,\sigma ^{2}\,e^{2\,\phi }\,\Delta ^{2},\\
{C_{2 \,3}} & = & {\displaystyle \frac {1}{2}} \,\Delta ^{3}\,\sigma \,e^{2\,\phi } - {\displaystyle \frac {1}{2
}} \,\Delta ^{3}\,r^{2}\,\phi '^{2}\,\sigma \,e^{2\,\phi } \\
& - & \sigma ^{3}\,e
^{2\,\phi }\,\Delta ^{2}  + {\displaystyle \frac {1}{2}} \,r\,\Delta '\,\sigma '\,e^{2\,\phi
}\,\Delta ^{2} - 2\,\phi '\,\Delta '\,\sigma \,
\Delta ^{2},\\
{A_{1 \,1}} & = &  - \sigma ^{5} + \sigma ^{3}\,\Delta,\\
{A_{1 \,3}} & = & {\displaystyle \frac {1}{2}} \,r^{2}\,\phi '^{2}\,\sigma ^{3}\,e^{2\,
\phi }\,\Delta ^{2} - {\displaystyle \frac {1}{2}} \,r\,\sigma '\,\sigma ^{2}\,e^{
2\,\phi }\,\Delta ^{2},\\
{A_{1 \,4}} & = &  - 3\,\Delta ^{3}\,\sigma ' + \sigma '\,\sigma ^{2}\,\Delta ^{2} + {\displaystyle
\frac {1}{2}} \,r^{2}\,\phi '\,\sigma ^{3}\,e^{2\,\phi }\,\Delta ^{2} \\
 & - & 4\,\Delta ^{3}\,\phi '\,\sigma  + 4\,\phi '\,\sigma ^{3}\,\Delta ^{2} ,\\
{A_{1 \,5}} & = &  - \sigma ^{3}\,\Delta ^{2} + \Delta ^{3
}\,\sigma ,\\
{B_{1 \,3}} & = &  - {\displaystyle \frac {1}{2}} \,r\,\sigma '\,\sigma ^{2}\,e^{
2\,\phi }\,\Delta  - 2\,\phi ''\,\sigma ^{3}\,\Delta  \\
 & + & {\displaystyle \frac {1}{2}} \,r^{2}\,\phi '^{2}\,\sigma ^{3}\,e^{2\,
\phi }\,\Delta   +  3\,\Delta ^{2}\,\phi ''\,\sigma  \\
 & - & 9\,\Delta ^{2}\,\phi '\,\sigma ' + 2\,\phi '\,\sigma '\,\sigma ^{2}\,
\Delta  \\
 & - & 6\,\Delta ^{2}\,\phi '^{2}\,\sigma  + 4\,\phi '^{2}\,\sigma ^{3}\,\Delta ,\\
{C_{1 \,3}} & = & \Delta ^{3}\,\phi '' - 2\,\Delta ^{3}\,\phi '^{2} - 3\,\phi ''\,\sigma ^{2}\,\Delta ^{2} \\
 & + & 2\,\phi '\,\sigma '\,\sigma \,\Delta^{2} - {\displaystyle \frac {1}{2}} \,r\,\sigma '\,\sigma \,e^{2\,\phi }\,\Delta
^{2}  \\
 & + &  {\displaystyle \frac {3}{4}} \,\Delta ^{2}\,r^{
2}\,\phi '^{2}\,\sigma ^{2}\,e^{2\,\phi }  +  6\,\phi'^{2}\,\sigma ^{2}\,\Delta ^{2},\\
{C_{1 \,4}} & = & \phi '\,
\sigma ^{2}\,\Delta ^{2} - {\displaystyle \frac {1}{4}} \,r
\,\sigma ^{2}\,e^{2\,\phi }\,\Delta ^{2} - 3\,\Delta ^{3}\,\phi '.\\
\end{eqnarray*}

\end{document}